\documentclass{emulateapj}
\usepackage{apjfonts}
\renewcommand{\AA}{\ensuremath{\mathrel{\hbox{\rlap{\hbox{A}} \kern-.11em \raise1.1ex \hbox{\tiny$^\circ$}\,}}}}

\newcommand{\helens}{HE\,0230$-$2130}
\newcommand{\mglens}{MG\,J0414$+$0534}
\newcommand{\rxninelens}{RX\,J0911$+$0551}
\newcommand{\sdsslens}{SDSS\,J0924$+$0219}
\newcommand{\sdsstenlens}{SDSS\,1004+4112}
\newcommand{\pglens}{PG\,1115$+$080}
\newcommand{\rxelevenlens}{RX\,J1131$-$1231}
\newcommand{\hlens}{H\,1413$+$117}
\newcommand{\blens}{B\,1422$+$231}
\newcommand{\wfilens}{WFI\,J2033$-$4723}
\newcommand{\qlens}{Q\,2237$+$0305}

\newcommand{\hubble}{\textit{Hubble Space Telescope}}
\newcommand{\chandra}{\textit{Chandra}}
\newcommand{\cxo}{\textit{Chandra X-ray Observatory}}
\newcommand{\hst}{\textit{HST}}

\newcommand{\ergcms}{\ensuremath{\mathrm{erg~cm}^{-2}~\mathrm{s}^{-1}}}

\newcommand{\err}[2]{\ensuremath{^{_{+#1}}_{^{-#2}}}}

\begin{document}
%%%%%%%%%%%%%%%%%%%%%%%%%%%%%%%%%%%%%%%%%%%%%%
\shorttitle{Flux Ratio Anomalies in Quadruply Lensed Quasars}
\shortauthors{Pooley et al.}
\slugcomment{accepted to ApJ}
%%%%%%%%%%%%%%%%%%%%%%%%%%%%%%%%%%%%%%%%%%%%%%
\title{X-Ray and Optical Flux Ratio Anomalies in Quadruply Lensed Quasars: I. Zooming in on Quasar Emission Regions\altaffilmark{1}}

\author{David~Pooley\altaffilmark{2,3}, Jeffrey~A.~Blackburne\altaffilmark{4}, Saul~Rappaport\altaffilmark{4}, \& Paul~L.~Schechter\altaffilmark{4} }

\altaffiltext{1}{Based on observations obtained with the Magellan Consortium's Clay Telescope.}
\altaffiltext{2}{University of California at Berkeley, Astronomy Department, 601 Campbell Hall, Berkeley, CA 94720; dave@astron.berkeley.edu}
\altaffiltext{3}{Chandra Fellow}
\altaffiltext{4}{Massachusetts Institute of Technology, Department of Physics and Kavli Institute for Astrophysics and Space Research, 70 Vassar St., Cambridge, MA 02139; jeffb@space.mit.edu, sar@mit.edu, schech@achernar.mit.edu}
%%%%%%%%%%%%%%%%%%%%%%%%%%%%%%%%%%%%%%%%%%%%%%
\begin{abstract}

X-ray and optical observations of quadruply lensed quasars can provide a microarcsecond probe of the lensed quasar, corresponding to scale sizes of $\sim 10^2-10^4$ gravitational radii of the central black hole. This high angular resolution is achieved by taking advantage of microlensing by stars in the lensing galaxy.  In this paper we utilize X-ray observations of ten lensed quasars recorded with the \cxo\ as well as corresponding optical data obtained with either the \hubble\ or ground-based optical telescopes.  These are analyzed in a systematic and uniform way with emphasis on the flux-ratio anomalies that are found relative to the predictions of smooth lens models.  A comparison of the flux ratio anomalies between the X-ray and optical bands allows us to conclude that the optical emission regions of the lensed quasars are typically larger than expected from basic thin disk models by factors of $\sim$3--30. 

\end{abstract}

%%%%%%%%%%%%%%%%%%%%%%%%%%%%%%%%%%%%%%%%%%%%%%

\section{Introduction}
\label{sec:intro}

In this paper we carry out a systematic, uniform study of ten quadruply gravitationally lensed quasars (hereafter ``quads'') for which one or more \cxo\ and optical images exist.  We show how such observations can probe the lensed quasar on angular scales of a few micro-arcseconds.

\setlength{\hoffset}{-10mm}
\renewcommand{\arraystretch}{1.25}
\begin{deluxetable*}{rlrcllllcl}[t!]
\tablewidth{0pt}
\tablecaption{X-ray Fluxes and Flux Ratios\label{tab:xray}}
\tablehead{
\multicolumn{3}{l}{Lensed Quasar} & & \multicolumn{4}{c}{Image Flux Ratios\tablenotemark{b}}&& \colhead{LM unabs.\ $F_\mathrm{0.5-8\,keV}$\tablenotemark{c}}\\ %[0.2ex]
\cline{1-3} \cline{5-8}\\[-2.5ex]
\colhead{ObsID\tablenotemark{a}}& \colhead{Date}& \colhead{Exp.\ (s)}& & \colhead{HS/HM}& \colhead{HS/LM}& \colhead{HM/LM}& \colhead{LS/LM}&& \colhead{($10^{-14}~\ergcms$)}}
\startdata   
\multicolumn{3}{l}{\helens}&&      $B/A$                & $B/C$                & $A/C$                & $D/C$                && $F_C$                \\ \cline{1-3}
 1642  & 2000 Oct 14.4 & 14\,764&& 0.44\err{0.08}{0.07} & 0.70\err{0.13}{0.12} & 1.6\err{0.2}{0.2}    & 0.45\err{0.08}{0.07} && 5.9\err{2.8}{2.0}    \\[2ex]

\multicolumn{3}{l}{\mglens}&&      $A_2/A_1$            & $A_2/B$              & $A_1/B$              & $C/B$                && $F_B$                \\ \cline{1-3}
 417   & 2000 Jan 13.7 &  6\,578&& 0.82\err{0.18}{0.15} & 1.9\err{0.4}{0.4}    & 2.3\err{0.5}{0.4}    & 0.25\err{0.07}{0.06} && 12\err{7}{5}         \\
 418   & 2000 Apr 2.9  &  7\,437&& 0.50\err{0.11}{0.10} & 1.3\err{0.3}{0.3}    & 2.6\err{0.4}{0.4}    & 0.63\err{0.11}{0.10} && 13\err{6}{4}         \\
 421   & 2000 Aug 16.9 &  7\,251&& 0.38\err{0.12}{0.10} & 0.96\err{0.38}{0.33} & 2.5\err{0.5}{0.4}    & 0.45\err{0.09}{0.07} && 15\err{6}{5}         \\
 422   & 2000 Nov 16.6 &  7\,504&& 0.67\err{0.13}{0.12} & 1.8\err{0.4}{0.4}    & 2.6\err{0.5}{0.4}    & 0.65\err{0.11}{0.10} && 14\err{9}{5}         \\
 1628  & 2001 Feb 5.1  &  9\,020&& 0.35\err{0.09}{0.08} & 0.89\err{0.28}{0.24} & 2.5\err{0.4}{0.3}    & 0.35\err{0.06}{0.05} && 16\err{7}{5}         \\
 3395  & 2001 Nov 9.3  & 28\,413&& 0.90\err{0.09}{0.08} & 1.8\err{0.2}{0.2}    & 1.9\err{0.2}{0.2}    & 0.53\err{0.05}{0.05} && 13\err{3}{3}         \\
 3419  & 2002 Jan 9.0  & 96\,664&& 0.61\err{0.04}{0.04} & 1.3\err{0.1}{0.1}    & 2.1\err{0.1}{0.1}    & 0.42\err{0.02}{0.02} && 14\err{4}{3}         \\[2ex]

\multicolumn{3}{l}{\rxninelens}&&  $A/B$                & $A/D$                & $B/D$                & $C/D$                && $F_D$                \\ \cline{1-3}
 419   & 1999 Nov 2.7  & 28\,795&& 2.7\err{1.3}{0.8}    & 3.4\err{0.6}{0.5}    & 1.3\err{0.4}{0.4}    & 0.35\err{0.13}{0.11} && 1.9\err{0.8}{0.6}    \\
 1629  & 2000 Oct 29.8 &  9\,826&& 27\err{500}{27}      & 4.6\err{1.6}{1.3}    & 0.17\err{0.90}{0.73} & 0.72\err{0.39}{0.28} && 1.8\err{0.5}{0.4}    \\[2ex]

\multicolumn{3}{l}{\sdsslens}&&    $D/A$                & $D/B$                & $A/B$                & $C/B$                && $F_B$                \\ \cline{1-3}
 5604  & 2005 Feb 24.0 & 17\,944&& 0.14\err{0.07}{0.06} & 0.45\err{0.25}{0.19} & 3.2\err{0.8}{0.6}    & 0.42\err{0.19}{0.14} && 1.2\err{0.8}{0.5}    \\[2ex]

%\multicolumn{3}{l}{\sdsstenlens}&& $A/B$                & $A/C$                & $B/C$                & $D/C$                && $F_C$                \\ \cline{1-3}
% 5794  & 2005 Jan 2.0  & 80\,087&& 0.77\err{0.03}{0.02} & 0.96\err{0.04}{0.04} & 1.24\err{0.05}{0.04} & 0.57\err{0.03}{0.03} && 13.5\err{1.05}{1.00} \\
% 5794  & 2005 Jan 2.0  & 80\,087&& 0.77\err{0.03}{0.03} & 0.93\err{0.04}{0.04} & 1.20\err{0.05}{0.05} & 0.55\err{0.03}{0.03} && 13.5\err{1.05}{1.00} \\[2ex]

\multicolumn{3}{l}{\pglens}&&      $A_2/A_1$            & $A_2/C$              & $A_1/C$              & $B/C$                && $F_C$                \\ \cline{1-3}
 363   & 2000 Jun 2.8  & 26\,492&& 0.16\err{0.03}{0.03} & 0.62\err{0.13}{0.12} & 3.9\err{0.3}{0.3}    & 1.1\err{0.1}{0.1}    && 6.8\err{1.1}{1.0}    \\
 1630  & 2000 Nov 3.3  &  9\,826&& 0.28\err{0.06}{0.06} & 1.2\err{0.3}{0.3}    & 4.4\err{0.6}{0.5}    & 1.0\err{0.2}{0.1}    && 8.3\err{2.9}{2.4}    \\[2ex]

\multicolumn{3}{l}{\rxelevenlens}&&$A/B$                & $A/C$                & $B/C$                & $D/C$                && $F_C$                \\ \cline{1-3}
 4814  & 2004 Apr 12.2 & 10\,047&& 0.10\err{0.01}{0.01} & 0.22\err{0.03}{0.02} & 2.2\err{0.1}{0.1}    & 0.30\err{0.03}{0.02} && 50\err{5}{5}         \\[2ex]

\multicolumn{3}{l}{\hlens}&&       $A/B$                & $A/C$                & $B/C$                & $D/C$                && $F_C$                 \\ \cline{1-3}
 930   & 2000 Apr 19.7 & 38\,185&& 1.8\err{0.5}{0.4}    & 4.0\err{1.4}{0.9}    & 2.2\err{0.9}{0.6}    & 1.2\err{0.5}{0.3}    && 2.7\err{0.7}{0.6}     \\
 5645  & 2005 Mar 30.1 & 88\,863&& 1.7\err{0.6}{0.4}    & 1.5\err{0.3}{0.2}    & 0.9\err{0.3}{0.2}    & 0.7\err{0.2}{0.2}    && 3.3\err{0.5}{0.5}     \\[2ex]

\multicolumn{3}{l}{\blens}&&       $B/A$                & $B/C$                 & $A/C$                & $D/C$                && $F_C$                \\ \cline{1-3}
 367   & 2000 Jun 1.6  & 28\,429&& 0.68\err{0.06}{0.05} & 1.1\err{0.1}{0.1}     & 1.6\err{0.1}{0.1}    & 0.11\err{0.01}{0.01} && 37\err{6}{5}         \\
 1631  & 2001 May 21.5 & 10\,651&& 0.62\err{0.09}{0.08} & 0.87\err{0.13}{0.12}  & 1.4\err{0.2}{0.2}    & 0.08\err{0.02}{0.02} && 40\err{12}{10}       \\
 4939  & 2004 Dec 1.6  & 47\,729&& 0.55\err{0.04}{0.04} & 0.95\err{0.08}{0.07}  & 1.7\err{0.1}{0.1}    & 0.10\err{0.01}{0.01} && 33\err{5}{5}         \\[2ex]

\multicolumn{3}{l}{\wfilens}&&     $A_2/A_1$            & $A_2/B$              & $A_1/B$              & $C/B$                && $F_B$                \\ \cline{1-3}
 5603  & 2005 Mar 10.1 & 15\,420&& 1.1\err{0.3}{0.2}    & 1.0\err{0.2}{0.2}    & 0.87\err{0.16}{0.14} & 0.64\err{0.11}{0.10} && 3.5\err{0.9}{0.8}    \\[2ex]

\multicolumn{3}{l}{\qlens}&&       $D/A$                 & $D/B$                & $A/B$                & $C/B$               && $F_B$                \\ \cline{1-3}
 431   & 2000 Sep 6.7  & 30\,287&& 0.17\err{0.01}{0.01}  & 0.85\err{0.09}{0.08} & 5.0\err{0.4}{0.3}    & 2.1\err{0.2}{0.2}   && 5.9\err{1.3}{1.1} \\
 1632  & 2001 Dec 8.8  &  9\,538&& 0.20\err{0.03}{0.03}  & 0.95\err{0.20}{0.17} & 4.7\err{0.7}{0.6}    & 1.7\err{0.3}{0.3}   && 6.1\err{3.7}{2.6} 
\enddata
%% 629,750 seconds of exposure
\tablenotetext{a}{The observation identifier of the \chandra\ dataset.}
\tablenotetext{b}{HS = Highly magnified Saddle point; HM = Highly magnified Minimum; LS = Less magnified Saddle point; LM = Less magnified Minimum. See \S\ref{sec:anomalies}.}
\tablenotetext{c}{The unabsorbed flux of the LM image is computed from the best fit power-law model described in \S\ref{sec:fx}.}
\end{deluxetable*}

Simple models for the gravitational potentials of lensing galaxies are usually sufficient to reproduce the positions of 4-image gravitationally lensed quasars.  But these same models -- a monopole plus a quadrupole -- fail to reproduce the optical fluxes of those images.  Such ``flux ratio anomalies'' are thought to be the product of small scale structure in the gravitational potentials of galaxies \citep{1995ApJ...443...18W, 1998MNRAS.295..587M, 2002ApJ...565...17C, 2001ApJ...563....9M, 2002ApJ...572...25D, 2002ApJ...580..685S}.  Of the two leading explanations, the more intriguing is that we are seeing the effects of dark matter condensations of sub-galactic mass.  The more prosaic explanation (though exciting for very different reasons) is that the anomalies are largely the result of microlensing by stars in the intervening galaxy.

A mass condensation can produce an anomaly only if the radius of its Einstein ring (a ``circle of influence'') is large compared to the emitting region of the lensed source.  To set the scale for the discussion, the Einstein radius of a star in a typical lensing galaxy is $\sim$$3\sqrt{(m/M_\odot)({\rm Gpc}/D_L)}$ microarcseconds, where $D_L$ is the angular diameter distance of the lens and $m$ is the mass of the star.  If the optical continuum emission from the typical quasar originated from something like a \citet{1973A&A....24..337S} thin accretion disk, we estimate that the Einstein ring of a 0.7 $M_\odot$ star would be much bigger than the optical emitting region. The stars in the lensing galaxy would then be expected to produce microlensing of the optical continuum \citep{2005ApJ...628..594M}.  As X-ray emission is generally thought to arise from an even smaller region than the optical continuum, the X-ray fluxes of quasar images ought likewise to exhibit such anomalies when observed with \chandra.

This expectation for the X-rays is borne out by observations. \cite{2001ApJ...555....1M} found that in the quadruple system \rxninelens\ the flux ratio of the $C$ image to the $D$ image was 0.19 in the X-ray compared with 1.28 in the optical.  \citet{2006ApJ...640..569B} found that in the quadruple system \rxelevenlens\ the $A/B$ ratio was 0.18 in the X-ray compared with 1.10 in the optical and model predictions in the vicinity of 1.70. More recent X-ray observations show considerable changes in the flux ratios of the images in this lens \citep{2006astro.ph..9112K}.  And \citet{2006ApJ...648...67P} find that in the quadruple system \pglens\ the $A_2/A_1$ ratio is roughly 0.16 in the X-ray compared with 0.68 in the optical and model predictions more nearly equal to unity.

In each of the above cases, the sense of the anomaly is that the X-ray flux ratios are yet more anomalous (in the sense of disagreeing with the models) than the optical flux ratios.  In previous work (cited above) we showed that this could happen only if the optical continuum emitting region were substantially larger than predicted for Shakura-Sunyaev disks.  In a number of individual studies  --- the three aforementioned cases, as well as \mglens\ \citep{2002ApJ...568..509C}, \hlens\ \citep{2004ApJ...606...78C}, and \qlens\ \citep{2003ApJ...589..100D} --- microlensing has been invoked to explain some of the observed X-ray properties.  In this paper we report the results of a systematic study of a larger sample of X-ray imaged quad lenses\footnote{The systems are the quadruple lenses \helens\ \citep{1999A&A...348L..41W}, \mglens\ \citep{1992AJ....104..968H}, \rxninelens\ \citep{1997A&A...317L..13B}, \sdsslens\ \citep{2003AJ....126..666I}, \pglens\ \citep{1980Natur.285..641W}, \rxelevenlens\ \citep{2003A&A...406L..43S}, \hlens\ \citep{1988Natur.334..325M}, \blens\ \citep{1992MNRAS.259P...1P}, \wfilens\ \citep{2004AJ....127.2617M}, and \qlens\ \citep{1985AJ.....90..691H}.}.   

The above discussion illustrates how microlensing permits at least some resolution of a quasar on microarcsecond scales, two orders of magnitude better than VLBI.  This corresponds to physical scales in the accretion disk of just a few thousand AU, or $\sim$1000 gravitational radii for a $\sim$$3\times10^8 M_\odot$ black hole at an angular diameter distance of 1 Gpc.

In addition, as suggested above, lensed systems present unique opportunities to study not only the lensed object but also the lensing object.  As we will discuss in a forthcoming paper, these same observations of X-ray flux ratio anomalies permit measurements of the dark matter content of the lensing galaxies \citep{2004IAUS..220..103S}.

In \S\ref{sec:xray-obs} we discuss the analysis of the \chandra\ archival data for ten quads.  In \S\ref{sec:opt-obs} we describe the properties of the selected group of lenses in the optical band.  We also present a uniform set of models for these lenses produced with the same software and for a common set of model assumptions.  The flux ratio anomalies are compared between the X-ray and optical images in \S\ref{sec:anomalies}.  In \S\ref{sec:quasars} we draw conclusions concerning the sizes of optically emitting regions in these ten sources.  Finally, in \S\ref{sec:conclusions} we summarize our findings.  Throughout this paper we adopt a cosmology with $\Omega_\Lambda = 0.7$, $\Omega_M = 0.3$, and $H_0 = 70$ km s$^{-1}$ Mpc$^{-1}$.

\begin{figure*}   
\centering
\includegraphics{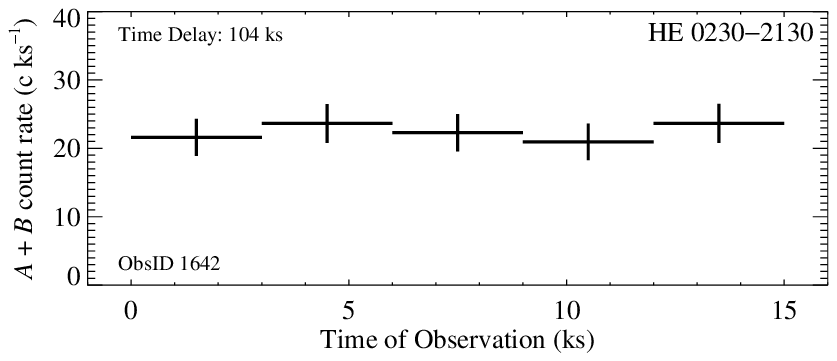}\hglue0.08in\includegraphics{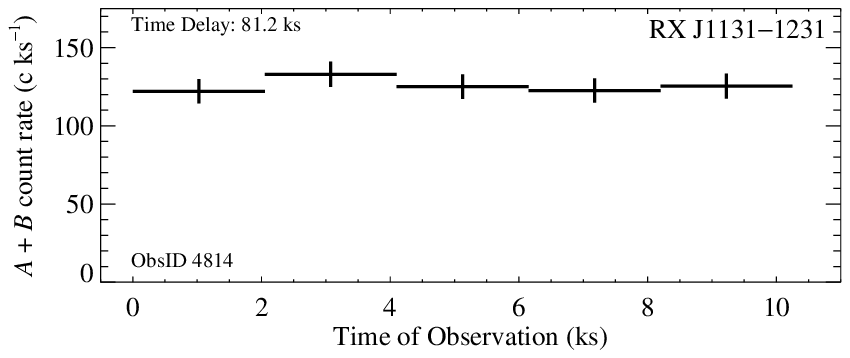}
\includegraphics{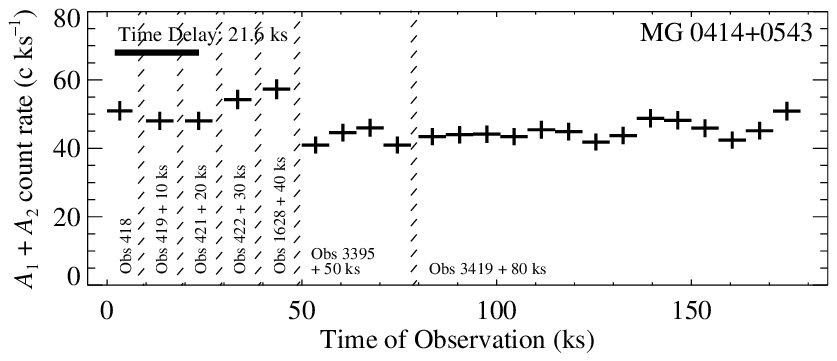}\hglue0.08in\includegraphics{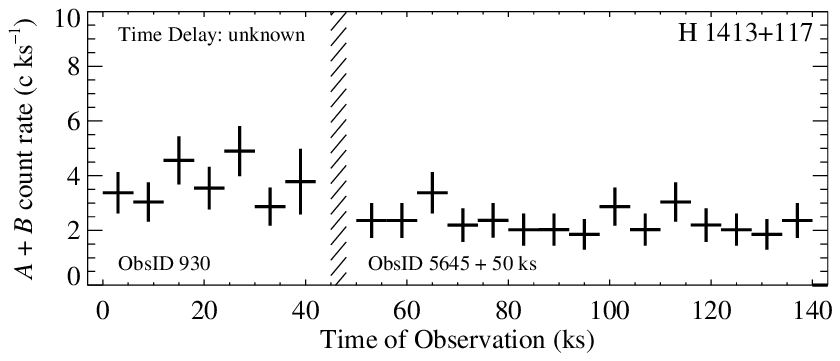}
\includegraphics{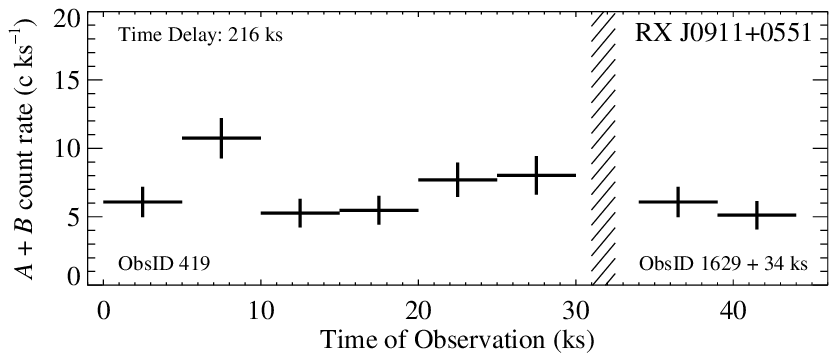}\hglue0.08in\includegraphics{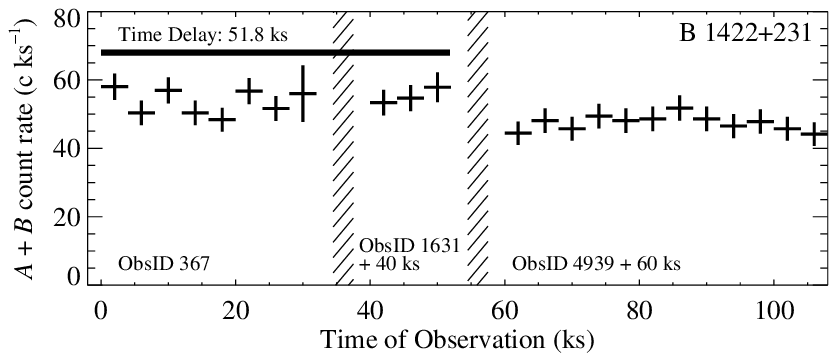}
\includegraphics{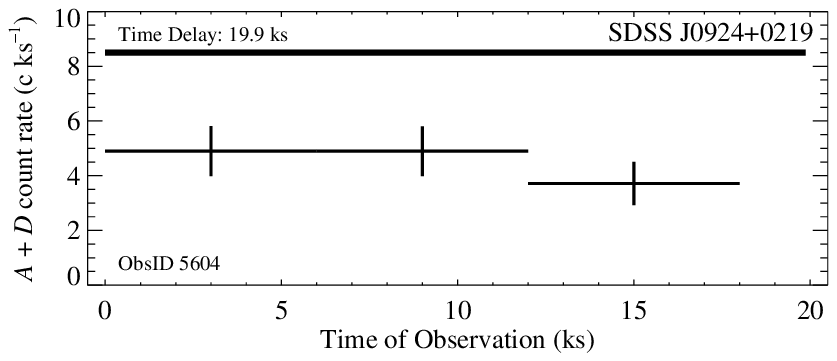}\hglue0.08in\includegraphics{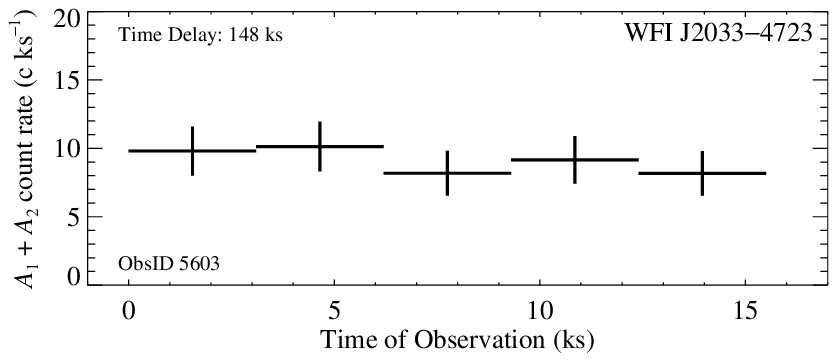}
\includegraphics{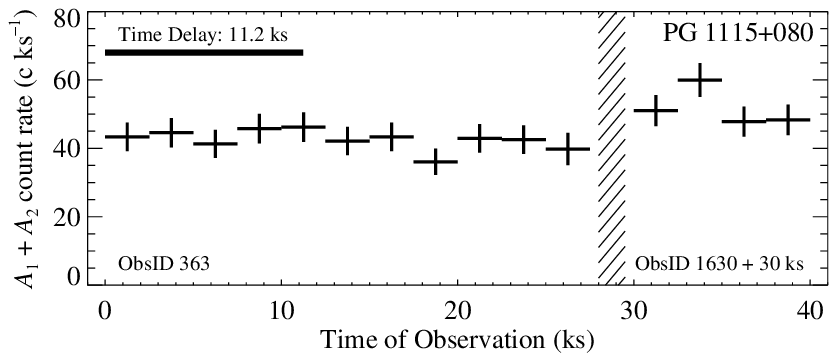}\hglue0.08in\includegraphics{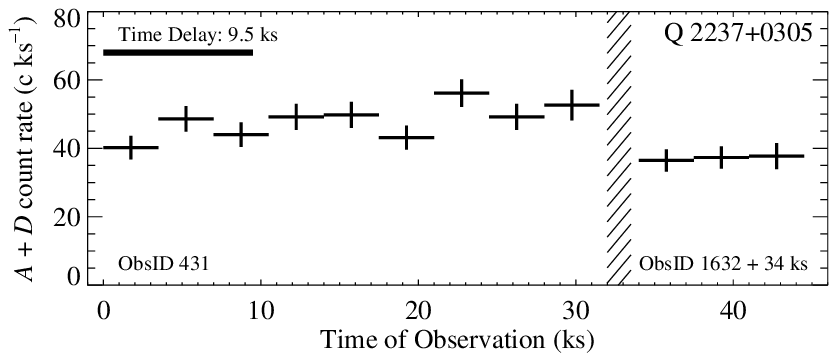}
\caption{X-ray lightcurves of the high magnification pair of images in each quad.  All observations in Table~\ref{tab:xray} were used to make these 0.5--8 keV lightcurves, with multiple observations of the same quad separated by hash marks.  The time delay between the pair (from the models described in \S\ref{sec:opt-obs} and Table~\ref{tab:lensmodels}) is given and shown as a thick horizontal bar in the cases where it will fit on the plots.}
\label{fig:xraylc}
\end{figure*}

%%%%%%%%%%%%%%%%%%%%%%%%%%%%%%%%%%%%%%%%%%%%%%
\section{X-ray observations}
\label{sec:xray-obs}

The data were downloaded from the \chandra\ archive, and reduction was performed using the CIAO\,3.3 software provided by the \chandra\ X-ray Center\footnote{\url{http://asc.harvard.edu}}.  The data were reprocessed using the CALDB\,3.2.2 set of calibration files (gain maps, quantum efficiency, quantum efficiency uniformity, effective area) including a new bad pixel list made with the {\tt acis\_run\_hotpix} tool.  The reprocessing was done without including the pixel randomization that is added during standard processing.  This omission slightly improves the point spread function.  The data were filtered using the standard {\it ASCA} grades and excluding both bad pixels and software-flagged cosmic ray events. Intervals of strong background flaring were searched for, and a few were found. In all cases, the flares were mild enough that removing the intervals would have decreased the signal-to-noise of the quasar images since it would have removed substantially more source flux than background flux within the small extraction regions.  Therefore, we did not remove any flaring intervals. The observation IDs, dates of observation, and exposure times are given in Table~\ref{tab:xray}.

\subsection{Determining X-ray Flux ratios}
\label{sec:fx}
For each observation, a sky image was produced in the 0.5--8 keV band with a sampling of 0\farcs0246 per pixel.  Because of the significant overlap of the lensed images (especially the close image pairs) in many cases, the intensities were determined by fitting to each sky image a two-dimensional model consisting of four Gaussian components plus a constant background.  The background component was fixed to a value determined from a source-free region near the lens.  The relative positions of the Gaussian components were fixed to the separations given in the CASTLES online database\footnote{\url{http://www.cfa.harvard.edu/castles/}}, but the absolute position was allowed to vary.  Each Gaussian was constrained to have the same full-width at half-maximum (FWHM), but this value was allowed to float.  The fits were performed using \citet{1979ApJ...228..939C} statistics and the Powell minimization method in Sherpa \citep{2001SPIE.4477...76F}.  

From the best fit 4-Gaussian model, the image flux ratios were calculated for the high magnification pair (saddle point and minimum; HS \& HM, respectively) as well as for each image relative to the less magnified minimum (LM) image.  The uncertainties on these ratios were determined with Sherpa via the {\tt projection} command, which varies each ratio in turn along a grid of values while all other parameters are allowed to float to the new best-fit values.  The results are given in Table~\ref{tab:xray}. 

Because the \chandra\ point-spread function is only approximately described by a Gaussian, we sought to test this method by utilizing a \chandra\ observation (ObsID 5794) of the large-separation quad \sdsstenlens, for which all four images are well separated\footnote{\sdsstenlens\ \citep{2003Natur.426..810I} is lensed by a dark matter-dominated cluster of galaxies, and is unique enough that we did not include it in our sample of lenses.}.  We extracted counts from the 90\% encircled energy region of each image, as determined by ACIS Extract v3.94 \citep{Broos02}, and formed a number of flux ratios. We also followed the above method of fitting Gaussians.  The agreement in flux ratios is excellent (see Table~\ref{tab:1004}).

\setlength{\hoffset}{0mm}
\begin{deluxetable}{lllll}[b]
\tablewidth{0pt}
\tablecaption{Comparison of Gaussian Fitting to Aperture Extraction of \sdsstenlens \label{tab:1004}}
\tablehead{& \multicolumn{4}{c}{X-ray Image Flux Ratios} \\ 
\cline{2-5}\\[-2.5ex]
\colhead{Method} & \colhead{$A/B$}& \colhead{$A/C$}& \colhead{$B/C$}& \colhead{$D/C$}}
\startdata
Gaussian fit & 0.77\err{0.03}{0.02} & 0.96\err{0.04}{0.04} & 1.24\err{0.05}{0.04} & 0.57\err{0.03}{0.03}\\
Extraction &   0.77\err{0.03}{0.03} & 0.93\err{0.04}{0.04} & 1.20\err{0.05}{0.05} & 0.55\err{0.03}{0.03}
\enddata
\end{deluxetable}

\begin{deluxetable}{llllll}
\tablewidth{0pt}
\tablecaption{Optical Observations
\label{tab:observations}}
\tablehead{\colhead{Lensed Quasar} & \multicolumn{4}{c}{Optical magnitudes} & \\ 
           \colhead{Obs.\ date}  & \colhead{HS} & \colhead{HM} & \colhead{LS} & \colhead{LM} &\colhead{Filter}}
\startdata
\helens    & $B$  & $A$  & $D$  & $C$  &\\
2002 Jul 29& 19.22& 19.02& 21.21& 19.59& F814W\tablenotemark{a}\\[2ex]
\mglens    & $A_2$& $A_1$& $C$  & $B$  &\\
1994 Nov 08& 21.36& 20.43& 22.10& 21.24& F814W\tablenotemark{a}\\[2ex]
\rxninelens& $A$  & $B$  & $C$  & $D$  & \\
2000 Mar 02& 18.38& 18.64& 19.36& 19.66& F814W\tablenotemark{a}\\[2ex]
\sdsslens  & $D$  & $A$  & $C$  & $B$  & \\
2003 Nov 19& 21.59& 18.69& 19.86& 19.52& F814W\tablenotemark{b}\\[2ex]
\pglens    & $A_2$  & $A_1$  & $B$ & $C$ & \\
2004 Feb 22& 15.86 & 16.08 & 17.68 & 17.26 & $i'$\tablenotemark{c}\\[2ex]
\rxelevenlens& $A$& $B$  & $D$  & $C$  & \\
2004 Apr 12& 17.43& 17.42& 19.72& 18.44& $I$\tablenotemark{d}\\[2ex]
\hlens     & $A$  & $B$  & $D$  & $C$  &\\
1994 Dec 22& 17.77& 17.84& 18.15& 18.06& F814W\tablenotemark{a}\\[2ex]
\blens     & $B$  & $A$  & $D$  & $C$  &\\
1999 Feb 06& 15.85& 15.88& 19.68& 16.41& F814W\tablenotemark{a}\\[2ex]
\wfilens   & $A_2$& $A_1$& $C$  & $B$  &\\
2003 Aug 01& 19.14& 18.68& 19.41& 19.32& $i'$\tablenotemark{e}\\[2ex]
\qlens     & $D$  & $A$  & $C$  & $B$  &\\
1999 Oct 20& 17.39& 15.92& 16.77& 17.21& F814W\tablenotemark{a}
\enddata
\tablenotetext{a}{See \url{http://www.cfa.harvard.edu/castles}}
\tablenotetext{b}{\citet{2006ApJ...639....1K}}
\tablenotetext{c}{Relative magnitudes from \citet{2006ApJ...648...67P}; zeropoint from this work.}
\tablenotetext{d}{\citet{2006AA...449..539S}}
\tablenotetext{e}{\citet{2004AJ....127.2617M}}
\end{deluxetable}

\begin{figure*}                                                                 
\centering
\includegraphics[width=\textwidth]{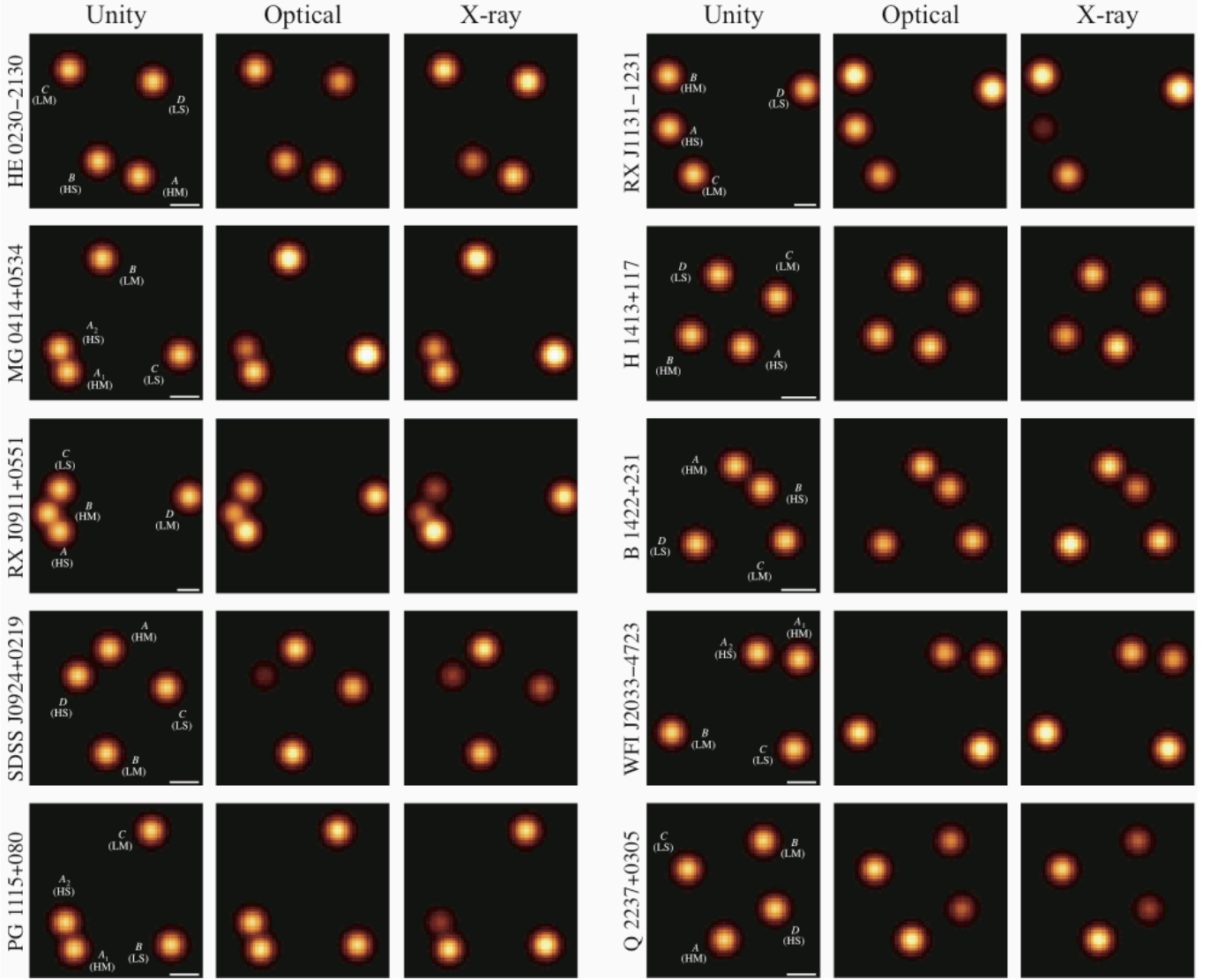}
\caption{Representation of the deviations from the models in X-rays and optical.  Each of the three frames for a system is constructed by placing Gaussians at the relative positions taken from the CASTLES online database.  The leftmost frame in each set has the intensity of each Gaussian set to unity.  In the center frame, the intensities are set to the ratio of the optical flux (normalized by the optical rms as defined by eq.\.(1)) to the model flux (normalized by the model rms; see eq.\,(1)).  The same is done for X-rays in the rightmost frame of each set. The same color scaling is applied to every frame. For aesthetic reasons, the FWHMs of the Gaussians are a constant fraction of the frame size; a 0\farcs5 scale bar is shown at the bottom right of each ``unity'' frame, and this frame also gives the image names and image types (see \S\ref{sec:anomalies}).}
\label{fig:images}
\end{figure*}

\begin{deluxetable*}{lccccccccc}[ht!]
\tablewidth{0pt}
\tablecaption{Lens Models
\label{tab:lensmodels}}
\tablehead{\colhead{} & \colhead{} & \colhead{} & \colhead{} & \multicolumn{4}{c}{Magnification\tablenotemark{b}} & \colhead{} & \colhead{}\\ \cline{5-8}\\[-2.5ex] \colhead{} & \colhead{$\theta_E$} & \colhead{$\gamma$} & \colhead{$\phi_{\gamma}$\tablenotemark{a}} & \colhead{HS} & \colhead{HM} & \colhead{LS} & \colhead{LM} & \colhead{$z_l$} & \colhead{$z_s$}}
\startdata
\helens\tablenotemark{c} & \nodata & \nodata &  \nodata & $-$11.80 & 11.50 & $-$2.32 & 6.22 & 0.52  & 2.162  \\
\mglens       & 1\farcs{}20 & 0.13 &  77.1$^{\mathrm o}$ & $-$20.72 & 19.07 &  $-$1.68 &  5.36 & 0.96  & 2.64  \\
\rxninelens   & 1\farcs{}11 & 0.32 &   1.1$^{\mathrm o}$ &  $-$4.41 &  8.10 &  $-$3.23 &  1.77 & 0.77  & 2.80  \\
\sdsslens     & 0\farcs{}88 & 0.04 &  84.6$^{\mathrm o}$ & $-$23.19 & 26.78 & $-$12.57 & 10.98 & 0.39  & 1.524 \\
\pglens       & 1\farcs{}15 & 0.12 &  65.0$^{\mathrm o}$ & $-$13.37 & 14.54 &  $-$3.02 &  3.88 & 0.31  & 1.72  \\
\rxelevenlens & 1\farcs{}78 & 0.12 & $-$73.3$^{\mathrm o}$ & $-$23.72 & 13.93 &  $-$1.58 & 13.40 & 0.295 & 0.658 \\
\hlens        & 0\farcs{}61 & 0.11 &  21.8$^{\mathrm o}$ &  $-$5.17 &  5.46 &  $-$3.32 &  5.05 & \nodata & 2.55 \\
\blens        & 0\farcs{}78 & 0.27 & $-$54.6$^{\mathrm o}$ & $-$12.04 &  8.86 &  $-$0.35 &  5.69 & 0.34  & 3.62 \\
\wfilens      & 1\farcs{}12 & 0.15 &  26.3$^{\mathrm o}$ &  $-$6.61 &  7.69 &  $-$2.20 &  3.17 & 0.66  & 1.66   \\
\qlens        & 0\farcs{}88 & 0.07 &  67.1$^{\mathrm o}$ &  $-$9.81 &  9.21 &  $-$5.32 &  7.93 & 0.04  & 1.69   \\
\enddata
\tablenotetext{a}{Measured in degrees East of North.}
\tablenotetext{b}{Negative magnifications signify saddle point images.}
\tablenotetext{c}{\helens\ has a unique mass model with an extra companion galaxy. See text for lens parameters.}
\end{deluxetable*}

Finally, a spectrum of the LM image was extracted for each observation with ACIS Extract and fit in Sherpa via a simple absorbed power law.  The absorption consisted of a fixed Galactic component \citep{1990ARA&A..28..215D} plus a variable component.  This simple model provided an acceptable fit in all cases, and the additional absorption component was usually consistent with zero.  The 0.5--8 keV flux of the unabsorbed power law is given in Table~\ref{tab:xray}.

\subsection{X-ray Variability}
\label{sec:xvar}
As the numbers in Table~\ref{tab:xray} indicate, many of the flux ratios vary to some degree for the quads that have been observed multiple times.  This may be due to varying degrees of microlensing or to normal quasar variability combined with time delays among the images.  In fact, variability plus a time delay could masquerade as a flux ratio anomaly.  Figure~\ref{fig:xraylc} shows the X-ray lightcurves of the sum of the high magnification pair of images for each system, which are seen to be fairly constant in all systems; only small amplitude (factor of two or less) variability is observed, even in cases when the length of the observation exceeds the predicted time delay between the brightest images\footnote{Predictions of time delays are subject to large uncertainties, especially for pairs of images with small separations. Indeed, the time delay for \rxelevenlens\ is now known to be greater than predicted by a factor of $\sim 12$ \citep{2006astro.ph..1523M}. This is mitigated by the fact that the optical and X-ray observations for this particular source were made on the same day.}.

For the rest of the analysis, we utilized the observation with the highest signal to noise ratio for the quads observed multiple times by \chandra. We chose not to average over multiple epochs in order to avoid averaging out variations due to changes in microlensing. We use ObsID 3419 for \mglens, ObsID 419 for \rxninelens, ObsID 363 for \pglens, ObsID 5645 for \hlens, ObsID 4939 for \blens, and ObsID 431 for \qlens.

%%%%%%%%%%%%%%%%%%%%%%%%%%%%%%%%%%%%%%%%%%%%%%%%%%%%

\section{Optical Images and Lens Models}
\label{sec:opt-obs}

We turned to the existing literature for optical data with which to compare our X-ray flux ratios. For each lens, we used data near $8000 \mbox{\AA}$, either Sloan $i'$, Cousins $I$, or \hst\ F814W. An effort was made to choose the observations closest in time to the deepest \chandra\ observation. The dates of the observations, along with the optical bandpasses and the image magnitudes, may be found in Table~\ref{tab:observations}. The images are arranged according to their magnifications and parity (see \S\ref{sec:anomalies}).
Under ideal circumstances, the X-ray and optical observations would have been made on the same day, in order to minimize systematic errors resulting from quasar variability and microlensing variability. But for most of these lenses, such contemporaneous observations have not been made. Three lenses have X-ray and optical observations separated by about 6 to 10 years, three by 2 to 4 years, and four by 15 months or less. One of these, \rxelevenlens, was observed in both bands on the same day \citep{2006AA...449..539S}. 

These delays between observations can add systematic uncertainty to the results. However, there are reasons to believe that their effect is not a strong one. The general lack of strong quasar variability seen in X-rays (see \S\ref{sec:xvar}), coupled with the limited success of campaigns to measure lens time delays (which rely on quasar variability), suggest that quasars do not often vary by the factors that would be required to explain the flux ratio anomalies. The fact that \rxelevenlens\ has an extremely strong discrepancy between X-ray and optical flux ratios despite simultaneous observations in both bands shows that time variability cannot fully explain the anomalous ratios.

We used Keeton's (2001) Lensmodel software, v1.06, to model each of the ten lenses as a singular isothermal sphere (SIS) with an external shear. This model has seven free parameters (lens strength, shear strength ($\gamma$) and direction ($\phi_\gamma$), and the positions of source and lens), making it overconstrained by the ten input measurements (the positions of four images and the lensing galaxy). The position measurements were obtained from the online CASTLES database. The observed fluxes of the lens images were not used as constraints. 

The models fit the image positions fairly well in all cases except that of \helens, where the position of the $D$ image is significantly altered by a second galaxy. Since this lens has an obvious strong perturbation from a companion lens galaxy, we added a second mass component to the model. Allowing its position and strength to vary, and using its measured position as a constraint, gave us eleven free parameters and twelve constraints. We found that a steeper projected profile than isothermal was required for this second mass component, so we modeled it as a circular power-law profile with an index of $-1.3$. This model allows a much better fit to the data, and predicts an Einstein ring radius of 0\farcs79 for the main lensing galaxy and 0\farcs42 for the perturber, and an external shear of 0.10 in a direction 60.1$^\circ$ west of north.

Parameters for the remaining lenses may be found in Table~\ref{tab:lensmodels}. The predicted magnifications may be expected to vary with different choices of lens models at the 10\% level \citep{2002ApJ...567L...5M}.

Our model, in which the quadrupole term of the gravitational potential arises from an external tide, gives larger magnifications (and therefore smaller bolometric luminosities) than would a model in which the quadrupole is due to the flattening of the lens galaxy. \citet{2003ApJ...589..688H} have argued that the high ratio of quadruply lensed quasars to doubly lensed quasars can be explained if most of the quadrupole is tidal in origin.

%%%%%%%%%%%%%%%%%%%%%%%%%%%%%%%%%%%%%%

\section{Comparison of anomalous flux ratios: \\ X-ray vs. optical}
\label{sec:anomalies}

Figure \ref{fig:images} provides a visual guide to the optical-to-model and X-ray-to-model flux ratios of each quad.  It shows representations of each system using two-dimensional Gaussians, the positions of which come from the CASTLES database.  As a point of reference, the leftmost frame for each quad shows Gaussians of unit amplitude.  The center frame represents the optical-to-model ratio of the images, normalized by each rms (described below).  The amplitude $A_i$ of image $i$ is given by
\begin{equation}
A_i = \frac{F_{\mathrm{opt},i}/\mathrm{rms}_\mathrm{opt}}{|\mu_i|/\mathrm{rms}_{|\mu|}}
\end{equation} 
where $i=1,2,3,4$, $F_{\mathrm{opt},i}$ is the (linear) optical flux of image $i$, and $\mu_i$ is the image magnification from Table~\ref{tab:lensmodels}. The right frame gives a similar representation for the X-rays.  The rms of the optical (and X-ray) observations is first computed as 
\begin{equation}
\sqrt{\frac{1}{4}\sum_{i=1,4} \left(F_{\mathrm{opt},i}\right)^2}~~~~.
\end{equation}
However, because the rms can be dominated by one highly anomalous image, we remove the largest deviator and then recompute the rms.  The largest deviator is defined as the image $i$ with the maximum value of $|\log_{10}(A_{i,\mathrm{opt}}) + \log_{10}(A_{i,\mathrm{x\mbox{-}ray}})|$.  This new rms is then used in eq.~(1) to compute the amplitudes, the values of which are given in Table~\ref{tab:fluxratios}.

In every case save one, the most anomalous image was the highly magnified saddle point image. This is not surprising, since \citet{2002ApJ...580..685S} have shown that microlensing is likely to affect high-magnification saddle points most strongly. In order to give the lenses a uniform treatment, we have classified the four images in each lens according to their magnifications and the local morphology of the travel-time surface. Henceforth in this paper, ``HS'' will designate the highly magnified saddle-point and ``HM'' the highly magnified minimum.  Likewise, ``LS'' will designate the less magnified saddle-point and ``LM'' the less magnified minimum.

\begin{figure*}[ht!]                                                                 
\centering
\includegraphics[height=2.3in]{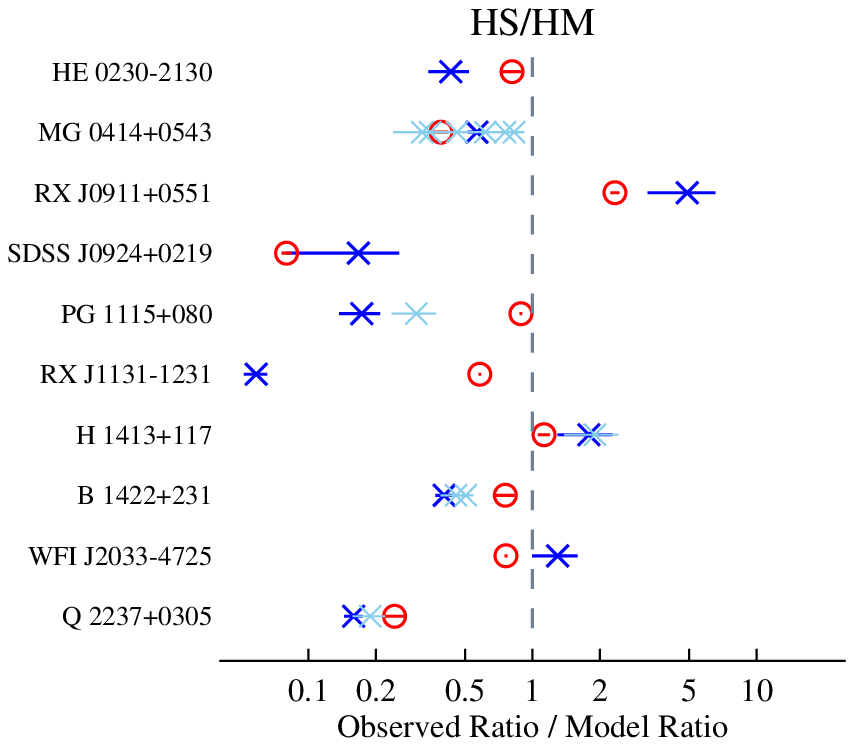}\hglue0.1in
\includegraphics[height=2.3in]{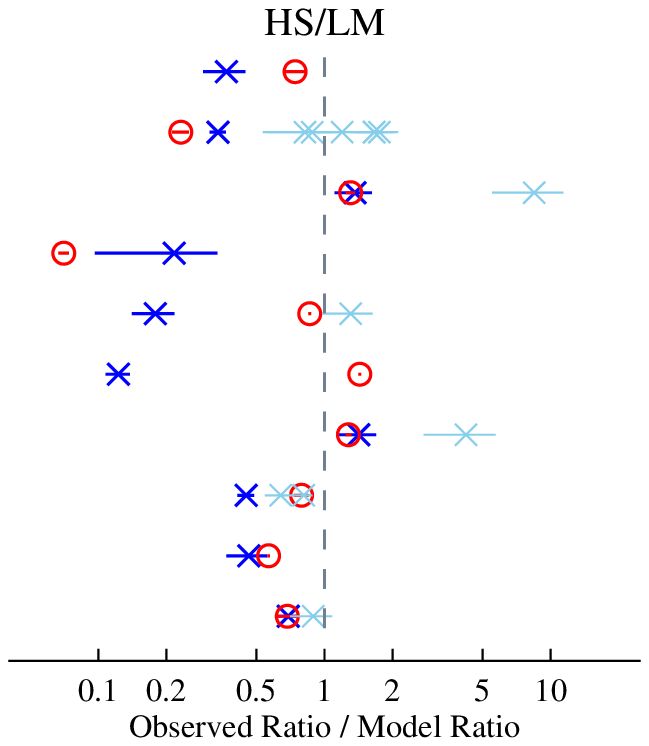}\hglue0.1in
\includegraphics[height=2.3in]{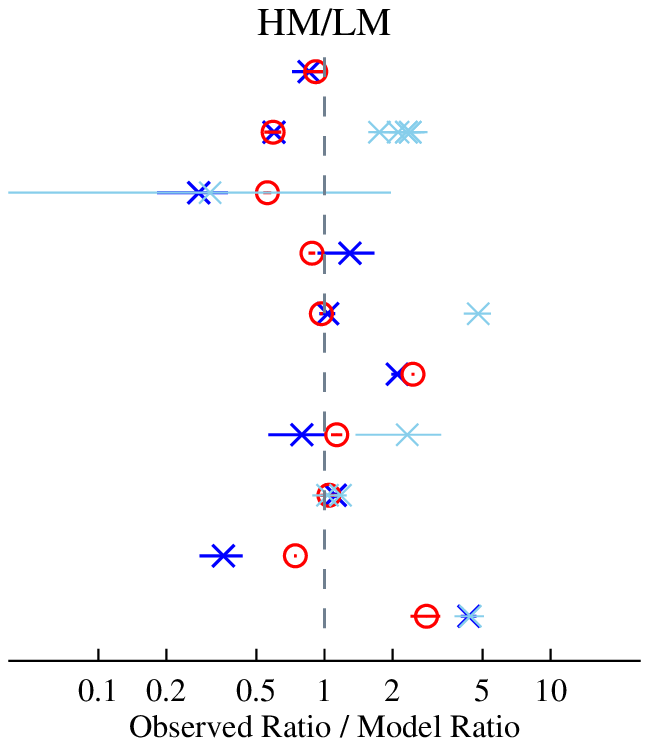}
\caption{Comparison of X-ray (blue {\boldmath$\times$}) and optical (red {\large\boldmath$\circ$}) ratios to lens model ratios for select image pairs for each lensed quasar.  The leftmost frame shows the ratio of the highly magnified saddle point (HS) to the highly magnified minimum (HM), while the center and rightmost frame show the ratio of each of these, respectively, to the less magnified minimum (LM).  The ratios for the X-ray are based on the observation with the highest signal to noise, and those for the optical are based on the observation closest in time to the chosen X-ray data.  The light blue {\boldmath$\times$}'s show the variation in the X-ray ratios for quads observed multiple times by \chandra.}
\label{fig:ratios}
\end{figure*}

In this work, we are most interested in the optical-to-model and X-ray-to-model ratios of the HS/HM flux ratio. The comparison between optical and X-ray ratios is shown for each quad in Figure~\ref{fig:ratios}.  The first panel shows the observed HS/HM ratio relative to the model HS/HM ratio, and the second and third panels show how each of HS and HM compare to the less magnified minimum image (LM).  In almost all cases, the HS/HM ratio is more extreme in X-rays than in the optical; when the observed ratio is greater then the model ratio, the X-ray ratio is greater than the optical, and, when the observed ratio is less than the model ratio, the X-ray ratio is less than the optical.  The second and third panels show whether the discrepancy with the model comes from the HS or the HM image (or a combination of the two).  In general, the LM image is much less susceptible to microlensing than either the HS or HM image \citep{2004ApJ...610...69K}.

The group statistics for the flux ratio anomalies presented in Figure~\ref{fig:ratios} and Table~\ref{tab:fluxratios} are summarized in Figure~\ref{fig:rms}. The error bars represent the $\pm$rms spread in the logarithm of the flux ratios (normalized by the smooth model values) between various image pairs for our quasar sample.  The black outer bars result from including all 10 quasars; the heavy blue bars result when we exclude the systems \qlens\ and \sdsslens. \qlens\ is excluded because the uniquely small redshift of its lensing galaxy causes the projected microlens Einstein radius to be bigger than any region of the source, while \sdsslens\ might also be excluded because the source size is thought to be so small that even its broad line region is partially microlensed \citep{2006ApJ...639....1K}.

It may be seen from the blue bars in this figure that the ratios of the HS to HM images deviate more (from their expected values) in the X-ray band than in the optical band by a factor of $\sim$2.4.  The discrepancy is somewhat smaller for the HS/LM ratios at a factor of $\sim$1.7.  The HM/LM and LS/LM ratios are not as anomalous in either band, but the X-ray ratios still have a wider range than do the optical ratios.  It is on the larger anomalies in the X-ray band for the HS/HM and HS/LM ratios, as compared to those for the optical band, that we base our quantitative analysis of the size of the optically emitting regions of the accretion disks in the next section.

%%%%%%%%%%%%%%%%%%%%%%%%%%%%%%%%%%%%%%%%%%%%%%

\section{Sizes of quasar emission regions}
\label{sec:quasars}

\begin{figure}[t]
\centering
\includegraphics[width=0.47\textwidth]{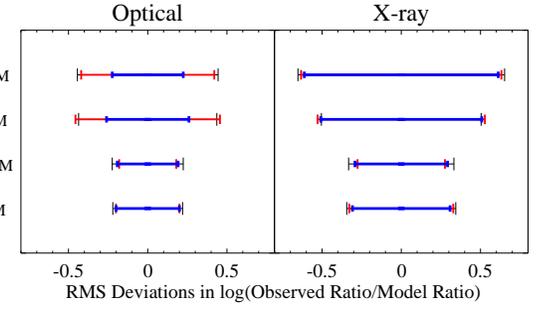} 
\caption{The rms of the flux ratio anomalies in the optical vs. X-ray (see Fig.\,\ref{fig:ratios} and Table 5 for the flux ratios of individual sources).  The black outer bars result from including all 10 quasars in our sample; the heavy blue bars result from excluding \qlens\ and \sdsslens; the red bars result when we exclude only \qlens.}
\label{fig:rms}
\end{figure}

For the purpose of interpreting our results, we adopt the working hypothesis that the anomalous flux ratios presented in this paper are the result of microlensing. Microlensing by stars in the lensing galaxy can account for the observed flux ratio anomalies, but only if the source is small compared to the Einstein radii of the microlensing stars. Figure~\ref{fig:ratios} shows dramatic evidence for microlensing in the X-ray band for at least 7 of the 10 lensing systems in our study.  In general, the optical emission of these same systems, while still being microlensed, has less extreme flux ratio anomalies than in the X-ray band by a factor of $\sim$2 (see Figure\,\ref{fig:rms} and the discussion above).  Since the X-rays are expected to be emitted very near to the black hole, the condition for microlensing is easy to meet -- the source should indeed be quite small compared to the Einstein radius of the microlensing stars.  By contrast, the markedly lower degree of microlensing in the optical band implies that the size of the optical emission region in many of these sources is roughly comparable to the size of the stellar microlens Einstein radius.

Many authors have studied the effect of source size on the microlensing of quasars by intervening galaxies.  Typically the results are presented as plots of microlensing light curves (e.g., Wambsganss \& Paczynski 1991) rather than rms fluctuations in the logarithm of the flux.\footnote{While the rms microlensing fluctuations in flux are formally divergent, rms fluctuations in the logarithm of the flux are not (e.g., Witt et al.\,1995).}  There are no analytic techniques for estimating rms fluctuations, so one must simulate the microlensing process.

Ideally we would run point source simulations for each of the 40 images in our sample, taking into account the theoretical magnification (which in turn depends upon two independent parameters, a convergence and a shear) and the fraction of baryonic matter.  Each simulation would produce a magnification map, which might then be convolved with sources of different sizes, producing magnification histograms.

Such an effort lies beyond the scope of the present paper, but we can draw upon such simulations that have been carried out.  In particular we use the work of Mortonson et al.\,(2005) who studied in detail the effect of source size on minima and saddlepoints with magnifications of $+6$ and $-6$ respectively, assuming that the convergence (a dimensionless surface density) is due entirely to equal mass stars and taking the convergence to be equal to the shear, as would be the case for an unperturbed isothermal sphere.  The magnifications observed for our highly magnified minima and saddlepoints are larger than this, typically by a factor of two, while our less magnified images are typically fainter than this by a factor of two.  Moreover there is reason to think that the stellar component comprises only a fraction -- somewhere between 1/10 and 1/2 -- of the mass surface density.  In the absence of a complete set of simulations we take those of Mortonson et al. as representative.

They find that, independent of the detailed radial profile of the source, the rms logarithmic fluctuations depend only upon the ratio of the half-light radius of the source to the Einstein radius, $r_{1/2}/r_\mathrm{Ein}$.  The rms logarithmic fluctuations decrease from their maximum value at $r_{1/2}/r_\mathrm{Ein} = 0$ to one half that value at $r_{1/2}/r_\mathrm{Ein} \approx 1/3$.  Since our optical fluctuations are roughly one half the amplitude of the X-ray fluctuations (which we take to arise from a region of negligible extent) we infer that the line-of-sight projected size of the optical region is roughly 1/3 the Einstein radius of the stars.

\begin{deluxetable}{llcccc}
\tablewidth{0pt}
\tablecaption{Flux-to-model Ratios Normalized by RMS
\label{tab:fluxratios}}
\tablehead{
\colhead{} & \colhead{} & \multicolumn{4}{c}{$\left(F_{\,i,\,\mathrm{obs}}/\mathrm{rms}_\mathrm{obs}\right) / \left(\left|\mu_i\right|/\mathrm{rms}_{\left|\mu\right|}\right)$} \\ \cline{3-6} \\[-2.5ex] 
\colhead{} & \colhead{Band} & \colhead{HS} & \colhead{HM} & \colhead{LS} & \colhead{LM}}
\startdata
\helens       & Optical & 0.76 & 0.94 & 0.62 & 1.02 \\
              &   X-ray & 0.46 & 1.06 & 1.50 & 1.24 \\[2ex]
\mglens       & Optical & 0.43 & 1.10 & 2.68 & 1.86 \\
              &   X-ray & 0.58 & 1.03 & 2.33 & 1.72 \\[2ex]
\rxninelens   & Optical & 1.49 & 0.64 & 0.82 & 1.14 \\
              &   X-ray & 1.76 & 0.36 & 0.24 & 1.29 \\[2ex]
\sdsslens     & Optical & 0.09 & 1.10 & 0.80 & 1.25 \\
              &   X-ray & 0.20 & 1.20 & 0.34 & 0.92 \\[2ex]
\pglens       & Optical & 1.02 & 1.15 & 1.04 & 1.19 \\
              &   X-ray & 0.20 & 1.14 & 1.48 & 1.10 \\[2ex]
\rxelevenlens & Optical & 1.03 & 1.76 & 1.87 & 0.72 \\
              &   X-ray & 0.10 & 1.72 & 2.08 & 0.82 \\[2ex]
\hlens        & Optical & 1.12 & 0.99 & 1.23 & 0.88 \\
              &   X-ray & 1.23 & 0.69 & 0.96 & 0.87 \\[2ex]
\blens        & Optical & 0.76 & 1.00 & 0.77 & 0.96 \\
              &   X-ray & 0.49 & 1.22 & 1.78 & 1.10 \\[2ex]
\wfilens      & Optical & 0.71 & 0.93 & 1.65 & 1.25 \\
              &   X-ray & 0.84 & 0.65 & 1.66 & 1.81 \\[2ex]
\qlens        & Optical & 0.33 & 1.36 & 1.08 & 0.48 \\
              &   X-ray & 0.22 & 1.41 & 1.00 & 0.33 \\
\enddata
\tablecomments{The rms values were computed from the three least anomalous images in each quad.  See \S\ref{sec:anomalies} for details.}
\end{deluxetable}

To estimate a rough size for the expected region of the optical emission from quasar accretion disks, we adopt a generic thin disk model \citep[see, e.g.,][]{1973A&A....24..337S}.  In such a model the gravitational energy release is redistributed via internal viscous stresses in such a way that, independent of the detailed nature of the origin of the viscosity, the rate of energy release per unit area of the disk at radius, $r$, is:
\begin{equation}
\mathcal{F} = \frac{3GM\dot M}{8\pi r^3} \left(1-\sqrt{\frac{r_0}{r}}\right)
\end{equation}
where $M$, $\dot M$, and $r_0$ are the black-hole mass, accretion rate, and the inner radius of the accretion disk, respectively. Note that in this formulation neither special nor general relativistic effects are included, except implicitly via the location of $r_0$. In our context, such relativistic effects are unimportant in the case of a Schwarzschild black hole.  Relativistic corrections, including those for accretion disks around Kerr black holes \citep[e.g.,][]{1974ApJ...191..499P} are only likely to exacerbate the difficulties with understanding the size of the optical emission regions discussed below.

\begin{figure}[t!]
\centering
\includegraphics[width=0.47\textwidth]{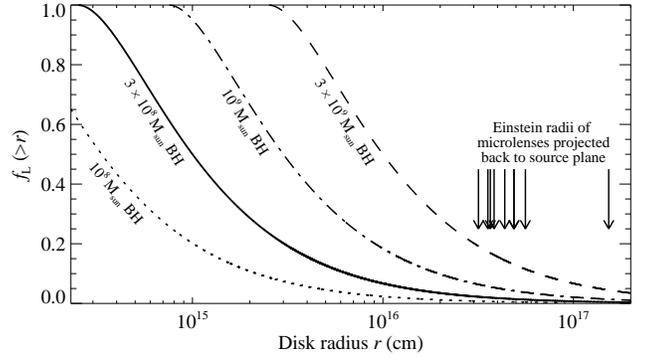}
\caption{Fraction of luminosity ($f_\mathrm{L}$) emitted beyond radius $r$ in a geometrically thin accretion disk for a variety of black hole masses.  The arrows indicate the physical sizes of the Einstein radii of $0.7\,M_\odot$ stars in each of the nine lensing galaxies of known redshift projected back onto the the lensed quasar.}
\label{fig:visc}
\end{figure}

In the context of the thin-disk model around a Schwarzschild black hole, the fractional luminosity that emerges within a radial distance $r$ is
\begin{equation}
f_L(<r) = 3r_0 \int_{r_0}^r \left(1-\sqrt{r_0/r}\right)r^{-2} dr = 1 - \frac{3}{r/r_0} + \frac{2}{\left(r/r_0\right)^{3/2}}~~.
\end{equation}
The complement of this quantity, $\{1-f_L(<r)\}$, i.e., the fraction of the luminosity released at radii $>r$, is plotted in Figure~\ref{fig:visc}. Here we have labeled the axis in physical units starting at $r_0=6GM/c^2 = 2.5 \times 10^{14}$ cm, i.e., the last stable orbit about a Schwarzschild black hole of $3 \times 10^8\,M_\odot$, an illustrative quasar mass.  We also show curves for other possible black hole masses.  For black holes with appreciable angular momentum, the value of $r_0$ moves progressively inward, and radii at which equal fractions of the luminosity are emitted do likewise.

Also overplotted on Figure~\ref{fig:visc} are nine arrows, one for each of our sources with known redshifts, marking the physical size of the Einstein radius of a $0.7\,M_\odot$ star in the lensing galaxy as projected back onto the lensed quasar.  What we see is that the arrows are virtually all located at radii where only a tiny fraction of the quasar luminosity can emerge from the disk -- at least for our fiducial black-hole mass of $3 \times 10^8\,M_\odot$. These fractional luminosity values are typically $\lesssim 2\%$ for sizes comparable to the backprojected stellar Einstein radii. Only for black-hole masses $\gtrsim 3 \times 10^9\,M_\odot$ does a significant fraction of the luminosity (i.e., $\sim$20\%) originate from radial distances comparable in size to the Einstein radius.  However, even then, as we showed in \citet{2006ApJ...648...67P} and demonstrate below, much of this radiation should be emitted at wavelengths well beyond the optical or near IR. Given that the optical radiation (e.g., 0.4--1.5 $\micron$) typically comprises a substantial fraction of quasar luminosities, e.g., $\sim$15\% \citep{1994ApJS...95....1E}, it appears difficult for the optical emission to be released from a thin disk at radii that are sufficiently large to allow for the partial suppression of microlensing --- as observed.  We further quantify this conclusion below.

Figure~\ref{fig:visc} and eq.\,(4) imply effective {\em upper limits} to the size of thin accretion disks in the optical by evaluating the {\em bolometric} luminosity emitted within a radial distance $r$ of the central black hole.  We now proceed to compute more quantitatively how large the accretion disk is expected to appear for a fixed waveband, e.g., $V$, $R$, $I$.  Based on the relativistic invariant $I_\nu/\nu^3$,we find the following expression for the half-light radius, $r_{1/2}$, of a thin accretion disk in  a waveband centered at $\nu$ (in the Earth's frame):
\begin{equation}
\frac{\int_{r_0}^{r_{1/2}}\left[e^{h\nu(1+z)/kT(r)}-1\right]^{-1}rdr}{\int_{r_0}^\infty \left[e^{h\nu(1+z)/kT(r)}-1\right]^{-1} rdr} = \frac{1}{2} ~~~,
\end{equation}
where $r_0$ is the location of the inner edge of the accretion disk, and $T(r)$ is the local temperature of the accretion disk, which in the Shakura \& Sunyaev (1973) model, is
\begin{equation}
T(r) = \left[\frac{3GM_{\rm BH} \dot M}{8\pi \sigma r^3} \right]^{1/4} \left( 1-\sqrt{r_0/r} \right)^{1/4} ~~~.
\end{equation}
In this simple picture, calculation of the half-light radius requires knowledge of three parameters: $M_{\rm BH}$, $\dot M$, and $r_0$.  We use primarily the optical-based method of \citet{2000ApJ...533..631K} (discussed below) to estimate the bolometric luminosity of each of the 10 sources in our sample.  We also utilize the X-ray luminosity, coupled with a bolometric correction factor (also discussed below) to provide a sanity check on the Kaspi et al.\,approach.  We further assume that all of the quasars are operating at the same fraction, $f_E \simeq1/4$,  of their respective Eddington limits \citep{2006ApJ...648..128K}.  We show below from a simple scaling argument, that our final results for $r_{1/2}$ are relatively insensitive to this choice.  Finally, we assume that the radiation efficiency (rest mass to radiant energy conversion efficiency, $\eta$) of all the quasars in our sample is $\eta = 0.15$ \citep[see, e.g., ][]{2002MNRAS.335..965Y}.  For this choice of efficiency, the dimensionless black-hole spin parameter would be $a=0.88$ and the innermost stable orbit would be located at $r_0 \simeq 2.5 R_g = 2.5 GM_{\rm BH}/c^2$ \citep[e.g., ][]{1970Natur.226...64B}.  However, in our simple non-relativistic disk model, we can only fix $r_0$, and accept whatever the non-relativistic energy release is.  For $r_0 =2.5\,R_g$ this turns out to yield an equivalent $\eta = 0.2$, which is sufficiently close to the Kerr value to provide the desired accuracy in computing $r_{1/2}$.

We summarize the computed and inferred properities of our quasar sample in Table \ref{tab:lum}.  The second column gives the bolometric luminosity as calculated from the \citet{2000ApJ...533..631K} prescription. In this approach, $L_{\rm bol}$ is taken to be $9 [\lambda F_{\lambda}]_{5100} 4 \pi d_L^2$. To estimate the 5100\,$\mbox{\AA}$ flux in the rest frame of the quasar, we used the flux measured in the closest available broadband filter, usually the \hst\ NICMOS F160W band, and extrapolated using an assumed power-law spectrum $F_{\lambda} \sim \lambda^{-1.7}$ \citep{2006ApJ...648..128K}. The third column in Table \ref{tab:lum} gives an independent estimate of the bolometric luminosity for each quasar based on the measured X-ray luminosity and a bolometric correction factor of 20, as inferred from the composite AGN spectrum of \citet{1994ApJS...95....1E}.  The fourth column provides the black-hole mass inferred from the bolometric luminosity (in column 2) divided by the Eddington fraction, $f_E=1/4$, which then yields $L_{\rm Edd}$, and thence $M_{\rm BH}$.  It should be noted that since Kollmeier et al.\,(2005) derive $f_E = 1/4$ using the prescription of \citet{2000ApJ...533..631K}, and since we follow suit, the masses we derive are independent of the dimensionless factor (of 9)  in Kaspi's prescription. The error bars on the mass represent the uncertainties inferred from the $\pm$ rms (logarithmic) spread in the bolometric luminosities obtained via three different estimates: (i) the \citet{2000ApJ...533..631K} method, (ii) the X-ray luminosity, and (iii) (in 7 of the 9 cases) the mass estimate directly provided by Peng et al.\,(1999; based on a ``virial'' method involving broad-line widths and sizes of broad-line regions). The values of the half-light radius, $r_{1/2}$, computed with eqs.\,(5) and (6) for the $I$ band are given in the fifth column.  In the sixth column are Einstein ring radii of typical $0.7 M_{\odot}$ microlenses, projected onto the plane of the source.  Finally, in the last column we give the logarithm of the ratio of the half-light radius to the microlens Einstein radius. 

\begin{figure}[t!]
\centering
\includegraphics[width=0.47\textwidth]{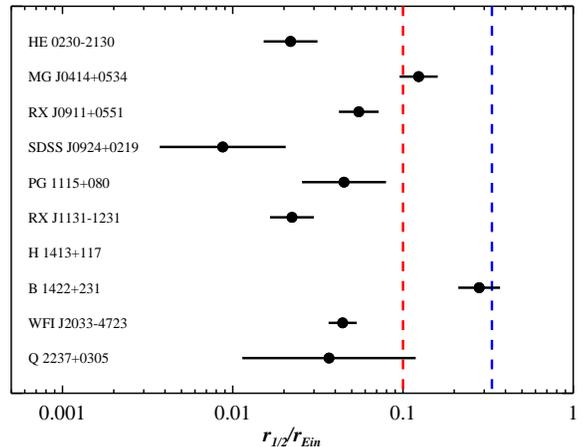}
\caption{Ratio of thin-disk half-light radii to typical (0.7 M$_{\odot}$) microlens Einstein radii for the ten sample lenses. Estimates of the minimum value required to significantly attenuate microlensing variability are 0.1 (red) and 0.33 (blue) -- see text for an explanation.}
\label{fig:rhrein}
\vglue0.2cm
\end{figure}

The results of our thin-disk estimates for the ratio $r_{1/2}/r_\mathrm{Ein}$ (last column of Table \ref{tab:lum}) are plotted in Figure\,\ref{fig:rhrein}.  The central heavy point within each error bar is based on the black-hole mass given in column 4 of Table \ref{tab:lum}.  The error bars on the ratio $r_{1/2}/r_\mathrm{Ein}$ are propagated from the black hole mass uncertainties given in Table \ref{tab:lum}.  The vertical line at $r_{1/2}/r_\mathrm{Ein} =1/3$ is our estimate of the ratio required to suppress microlensing in the optical band by the (logarithmic) factor of $\sim$2 discussed in \S\ref{sec:anomalies}.  The vertical line at $r_{1/2}/r_\mathrm{Ein}=1/10$ represents a more conservative lower limit on $r_{1/2}/r_\mathrm{Ein}$ that might plausibly still be consistent with the suppressed microlensing in the optical. An inspection of Figure\,\ref{fig:rhrein} shows that 7 of the 9 systems (for which $r_{1/2}/r_\mathrm{Ein}$ could be calculated) lie below the limit of 1/10, and therefore the disk size in the optical that we calculate appears to be too small to explain the reduced microlensing.  Only one of the systems, \blens, has a ratio of $r_{1/2}/r_\mathrm{Ein}$ that slightly exceeds the 1/3 value which we think is reasonable to account for the reduced microlensing in the optical for this particular source.  

\begin{deluxetable*}{lccccccc}[ht!]
\tablewidth{0pt}
\tablecaption{Quasar Properties
\label{tab:lum}}
\tablehead{\colhead{Quasar} & \colhead{$L_\mathrm{bol,opt}$\tablenotemark{a}} & \colhead{$L_\mathrm{bol,X}$\tablenotemark{b}} & \colhead{log $M_{\rm BH}$\tablenotemark{c}} & \colhead{$r_{1/2}$\tablenotemark{d}} & \colhead{$r_{1/2}$\tablenotemark{d}} & \colhead{stellar $r_\mathrm{Ein}$\tablenotemark{e}} & \colhead{log $r_{1/2}/r_\mathrm{Ein}$}\\
\colhead{} & \colhead{($10^{45}$ erg~s$^{-1}$)} & \colhead{($10^{45}$ erg~s$^{-1}$)} & \colhead{($M_\odot$)} & \colhead{($10^{15}$ cm)} & \colhead{($R_g$)} & \colhead{($10^{15}$ cm)} & \colhead{}}
\startdata
\helens       & 2.9 & 6.3  & $7.95 \pm 0.24$ & 0.93 & 70 &  43  & $-1.66 \pm 0.16$ \\
\mglens       & 36  & 28   & $9.04 \pm 0.17$ & 3.8  & 23 &  31  & $-0.91 \pm 0.11$ \\
\rxninelens   & 13  & 13   & $8.60 \pm 0.18$ & 1.9  & 32 &  35  & $-1.26 \pm 0.12$ \\
\sdsslens     & 0.6 & 0.3  & $7.27 \pm 0.56$ & 0.42 & 152&  48  & $-2.06 \pm 0.37$ \\
\pglens       & 11  & 6.6  & $8.53 \pm 0.37$ & 2.5  & 50 &  55  & $-1.35 \pm 0.25$ \\
\rxelevenlens & 0.80& 1.3  & $7.39 \pm 0.19$ & 0.84 & 230&  38  & $-1.65 \pm 0.13$ \\
\hlens        & 56  & 6.5  & $9.24 \pm 0.51$ & 5.4  & \nodata & \nodata & \nodata  \\
\blens        & 250 & 135  & $9.89 \pm 0.18$ & 13   & 11 &  47  & $-0.55 \pm 0.12$ \\
\wfilens      & 5.7 & 3.8  & $8.24 \pm 0.12$ & 1.6  & 62 &  36  & $-1.35 \pm 0.08$ \\
\qlens        & 32  & 2.7  & $8.99 \pm 0.76$ & 5.5  & 38 & 150  & $-1.43 \pm 0.51$ \\
\enddata
\tablenotetext{a}{Bolometric luminosities computed using $L_{\rm bol}=9 [\lambda F_{\lambda}]_{5100}4\pi d_L^2$. Computed from HM, LS, and LM images, corrected for magnification \citep{2000ApJ...533..631K}.}
\tablenotetext{b}{Approximate bolometric luminosities derived from the X-ray (0.5--8 keV) luminosities (computed from LM image) with a bolometric correction factor of 20 (see \S\ref{sec:quasars}).}
\tablenotetext{c}{Calculated from the bolometric luminosities in column 2. See \S\ref{sec:quasars}.}
\tablenotetext{d}{$r_{1/2}$ is computed according to eq.\,(5) for the I band.}
\tablenotetext{e}{Einstein radius of a $0.7 M_{\odot}$ star, projected back to the lensed quasar, in units of $10^{15}$ cm.}
\end{deluxetable*}

Arguably the largest contribution to the uncertainty in the calculation of $r_{1/2}$ arises from the errors in estimating the bolometric luminosities of the quasars in our sample.  We believe we can make a fairly robust estimate of the uncertainty in $r_{1/2}$ --- due to the uncertainties in $L_{\rm bol}$ --- by inspection of Figure\,\ref{fig:rhrein}.  As discussed above, the plotted error bars are derived from the logarithmic rms scatter among the three (two) different and independent methods we have employed to infer $L_{\rm bol}$ (see above discussion) for 7 (2) of the sources.  As Figure\,\ref{fig:rhrein} indicates, the uncertainties in $r_{1/2}$ range between factors of 1.2 and 3.2, with an average value of a factor of 1.7.  We take this to be a fairly reliable estimate of the uncertainty in our values for $r_{1/2}$ due to errors in estimating $L_{\rm bol}$.                           
 
In our calculations leading to the set of values for $r_{1/2}/r_\mathrm{Ein}$ we assumed values for two key parameters of the quasars: (i) the radiative efficiency $\eta$, and (ii) the fraction $f_E \equiv L_{\rm bol}/L_{\rm Edd}$.  Based on a simple scaling argument, we can show how our results for $r_{1/2}$ depend on $\eta$ and $f_E$. Equation (5) provides the exact definition of $r_{1/2}$ that we use.  However, if we use the expression for $T(r)$ in eq.\,(6) to find the radius where $T/(1+z)$ equals $h\nu/k$, where $\nu$ is the center of the observation band, this is to a good approximation proportional to $r_{1/2}$.  If we further neglect the factor $(1-\sqrt{r_0/r})$ in eq.\,(6), we find a handy scaling relation for $r_{1/2}$:
\begin{equation}
r_{1/2} \propto \left(M_{\rm BH} \dot M\right)^{1/3} ~~~.
\end{equation}
If we consider the bolometric luminosity to be a measured quantity for each system, then $M_{\rm BH} \propto L_{\rm bol}/f_E$ and $\dot M \propto L_{\rm bol}/\eta$.  Combining these, we can see how $r_{1/2}$ depends on the assumed parameters $f_E$ and $\eta$:
\begin{equation}
r_{1/2} \propto \left( f_E \eta \right)^{-1/3} ~~~,
\end{equation}
which is a fairly weak dependence, and not likely to lead to uncertainties in $r_{1/2}$ of more than an additional factor of $\sim$2.

%%%%%%%%%%%%%%%%%%%%%%%%%%%%%%%%%%%%%%%%%%%%%

\section{Summary and Conclusions}
\label{sec:conclusions}

We have presented a study of ten quadruply gravitationally lensed quasars for which high spatial resolution X-ray and optical data are available, paying particular attention to the differences between the observed flux ratios of the high magnification pairs of images (i.e., HS/HM) and the predicted flux ratios from smooth lensing models.  The \chandra\ data were analyzed in a uniform and systematic manner, and the X-ray flux ratios were determined via two-dimensional Gaussian fits.  The optical fluxes and image positions were found in the existing literature, with the bulk coming from the CASTLES project.  We also modeled each lensing system as a singular isothermal sphere with external shear (except for \helens, where a second mass component was necessary), and these simple models fit the image positions quite well.

As illustrated in Figures~\ref{fig:images} -- \ref{fig:rms}, almost all systems show evidence for an anomaly in the ratio of high-magnification saddle point and minimum images (HS/HM) as compared to the smooth model prediction.  In the systems which show a pronounced anomaly, the X-rays are generally seen to be more anomalous than the optical.

For a number of reasons, we believe that the anomalous flux ratios, and the differences between these ratios in the X-ray and optical bands, are best explained by microlensing.  In previous work \citep{2006ApJ...640..569B,2006ApJ...648...67P} we have shown that extinction in the visible band and absorption of soft X-rays cannot provide the explanation. Second, we show in this study (as well as previous work) that temporal variability intrinsic to the source, in conjunction with lens time delays, also cannot, in most cases, explain the observed anomalies.  Third, since images in both the X-ray and optical bands exhibit these flux ratio anomalies, but to differing degrees, no smooth lens model can reproduce these anomalies. Finally, we find that in the preponderance of systems, it is the highly magnified saddle point image (HS) whose flux is anomalous.  This is in agreement with microlensing magnification distributions \citep{2002ApJ...580..685S}.  Since there is no reason for the HS location to systematically produce larger optical extinctions or X-ray absorptions, this is another argument against differential extinction/absorption being the cause of the flux ratio anomalies.   

Under the hypothesis that the anomalies are produced via microlensing by stars (of typical mass 0.7~$M_\sun$) in the lensing galaxy, the implication is that the optical emitting region, which suffers rms (logarithmic) microlensing variations only half as big as those of the X-ray region, must have a typical size $\sim$1/3 of the Einstein radius of the microlensing stars (see discussion in \S\ref{sec:quasars}).  Likewise, the X-ray emitting region, being more severely microlensed, must be substantially smaller than this.

In the context of a thin accretion disk around a black hole, the X-ray requirement is easily satisfied, as this emission likely arises from the inner parts of the disk.  However, the optical emission poses something of a problem.  It is generally thought to arise from a region not much larger than the X-ray region, but this is in conflict with the observed microlensing results which require larger optical emitting regions by factors of $\sim 3-30$ (see Figure\,\ref{fig:rhrein}) than are commonly accepted.

Therefore, we are left with a conundrum.  Either there is a mechanism to transport the optical radiation to larger radii (and which does not affect the X-rays), or there is a missing piece of the puzzle.  Regardless, we have demonstrated how the X-ray and optical observations can provide a micro-arcsecond probe of the lensed quasars, and thereby yield potentially important results.   

From the work in this paper and the above discussion we draw three summary conclusions: \\
\indent$\bullet$ microlensing is the primary cause of the flux ratio anomalies. \\ 
\indent$\bullet$ the optical emitting regions in the quasars involved in this study have sizes of $\sim$1/3 of a stellar Einstein radius, i.e., $\sim$ a microacrsecond, corresponding to $\sim$1000 AU.\\
\indent$\bullet$ millilensing (e.g., by dark matter haloes) is ruled out as an explanation of the flux ratio anomalies by virtue of the above conclusion since this implies that both the X-ray and optical emission regions are small compared to the milliarcsecond scale, and should therefore be lensed by the same amount. 

Finally, as mentioned earlier in the paper, these same flux ratio anomalies can be used to provide valuable information on the ratio of stellar matter to dark matter in the lens galaxy in the vicinity of its Einstein radius. \citet{2004IAUS..220..103S} used {\it optical} flux ratio anomalies for a sample of eleven quads to derive a projected stellar/dark mass ratio at the typical impact parameter of a quasar image.  They first assumed that the optical emission region was small compared to the Einstein rings of the stars in the lensing galaxy.  The result was very heavily influenced by the inclusion (or exclusion) of the system with the most extreme flux ratio anomaly, \sdsslens.  They found less discordant results if they instead assumed that half the optical light came from a pointlike source and half came from a more extended source.  In this case the results indicated a stellar to dark matter ratio of $\sim$1/4 (at the typical impact parameter of a quasar image), with an uncertainty of a factor of about 2. But allowing for a fraction of the light to come from a more extended region adds a second parameter to the problem, making determination of the stellar/dark matter ratio more uncertain.  If, as we have argued here, the X-ray emission comes from a region substantially smaller than the optical emission region, the use of X-ray flux ratio anomalies in the analysis of Schechter \& Wambsganss would eliminate that second parameter and more uniquely determine the ratio of stellar matter to dark matter. The same study would give a better idea of the emission region size required to attenuate microlensing variations. Such an analysis is the subject of a forthcoming paper.

\acknowledgements 

We thank the anonymous referee for useful suggestions and Julian Krolik for extensive and very helpful discussions about this work.  We are grateful to Josiah Schwab for helping with the numerical accretion-disk calculations.  D.~P.\ gratefully acknowledges the support provided by NASA through \chandra\ Postdoctoral Fellowship grant PF4-50035 awarded by the \chandra\ X-Ray Center, which is operated by the Smithsonian Astrophysical Observatory for NASA under contract NAS8-03060.  S.~R.\ received some support from \chandra\ Grant TM5-6003X. J.~A.~B.\ and P.~L.~S.\ acknowledge support from NSF Grant AST-0206010.

\end{document}